\newcommand{\CLASSINPUTbottomtextmargin}{0.8in}
\documentclass[conference]{IEEEtran}
\usepackage{amsmath,amsfonts}
\usepackage{graphicx,amssymb}
\usepackage{algorithmic,algorithm}
\usepackage{amsthm,dsfont}
\usepackage{subfigure,url}
\usepackage[noadjust]{cite}

\theoremstyle{plain}
\newtheorem{thm}{Theorem}
\newtheorem{coro}[]{Corollary}
\newtheorem{prop}[]{Proposition}

\newcounter{longequ}[longequ]
\addtocounter{longequ}{8}

\newcommand\relphantom[1]{\mathrel{\phantom{#1}}}
\allowdisplaybreaks

\IEEEoverridecommandlockouts
\begin{document}
%
\title{A Tractable Framework for Coverage Analysis of Cellular-Connected UAV Networks}
\author{\IEEEauthorblockN{Xianghao Yu\IEEEauthorrefmark{1}, Jun Zhang\IEEEauthorrefmark{2}, Robert Schober\IEEEauthorrefmark{1}, and Khaled B. Letaief\IEEEauthorrefmark{3}}
\IEEEauthorblockA{\IEEEauthorrefmark{1}Friedrich-Alexander-Universit\"{a}t Erlangen-N\"{u}rnberg, Germany,
\IEEEauthorrefmark{2}Department of EIE, The Hong Kong Polytechnic University\\
\IEEEauthorrefmark{3}Department of ECE, The Hong Kong University of Science and Technology\\
Email: \IEEEauthorrefmark{1}\{xianghao.yu, robert.schober\}@fau.de, \IEEEauthorrefmark{2}jun-eie.zhang@polyu.edu.hk, \IEEEauthorrefmark{3}eekhaled@ust.hk}
}


%


\maketitle

\begin{abstract}
Unmanned aerial vehicles (UAVs) have recently found abundant applications in the public and civil domains.
To ensure reliable control and navigation, connecting UAVs to controllers via existing cellular network infrastructure, i.e., ground base stations (GBSs), has been proposed as a promising solution. 
Nevertheless, it is highly challenging to characterize the communication performance of cellular-connected UAVs, due to their unique propagation conditions. 
This paper proposes a tractable framework for the coverage analysis of cellular-connected UAV networks, which consists of a new blockage model and an effective approach to handle general fading channels. In particular, a \emph{line-of-sight (LoS) ball} model is proposed to capture the probabilistic propagation in UAV communication systems, and a tractable expression is derived for the Laplace transform of the aggregate interference with general Nakagami fading.
This framework leads to a tractable expression for the coverage probability, which in turn helps to investigate the impact of the GBS density.
Specifically, a tight lower bound on the optimal density that maximizes the coverage probability is derived.
Numerical results show that the proposed LoS ball model is accurate, and the optimal GBS density decreases when the UAV altitude increases.

\end{abstract}

\IEEEpeerreviewmaketitle

\section{Introduction}
Unmanned aerial vehicles (UAVs), also known as drones, have become prevalent in recent years and are getting integrated into the daily operations across industries. 
Reliable and long-range connectivity with ground pilots through wireless communications is of great importance to provide stable control, monitoring, and navigation for UAVs \cite{8438489}.
However, current UAV systems mainly operate in the unlicensed spectrum bands, suffering from low data rate, short operation range, and vulnerability to interference.

Given its wide-area coverage, as well as its scalability, security, and reliability, the cellular network stands out as a promising candidate for connecting UAVs with the ground pilots \cite{8337920}. 
In 2017, the 3rd Generation Partnership Project (3GPP) launched a project on the enhanced Long Term Evolution (LTE) support for UAVs \cite{instance1290}.
By integrating UAVs as new aerial users into the current LTE systems or the 5G networks of the near future, it is anticipated that significant performance improvement can be achieved over existing systems relying on unlicensed bands \cite{8470897}.
Therefore, evaluating the performance of cellular-connected UAV networks is a pivotal research task.
As field trials are costly and system-level simulations are time-consuming, an effective alternative is to quantify the network performance analytically.

There are three distinctive characteristics of cellular-connected UAVs compared to conventional terrestrial users. First, taking the UAV altitude into consideration, the link distances between transceivers are determined by 3-D positions, compared with the planar 2-D scenario in conventional terrestrial cellular networks.  
Second, different from conventional cellular networks where the transmissions mainly rely on non-line-of-sight (NLoS) multi-path scattering, cellular-connected UAVs often can establish links with LoS paths. This is because UAVs in the sky can avoid the nearby blockages on the ground, such as buildings and vegetation. 
Finally, the small-scale fading model describing LoS-dominated links is no longer Rayleigh fading as in conventional cellular networks with rich scattering. These key aspects complicate the network analysis of cellular-connected UAV networks, and call for new investigations.

There exist several previous studies on performance analysis of wireless networks incorporating UAVs \cite{7412759,7967745,8254658,azari2018cellular}. Initial studies were conducted in wireless networks where UAVs serve as mobile aerial base stations (BSs). The coverage probabilities were derived for terrestrial users by neglecting the small-scale fading in \cite{7412759}. In addition, Nakagami fading was assumed in \cite{7967745}. However, all 3-D links in the network were assumed to be LoS, which weakened the applicability of the obtained results. To address this problem, a measurement-based probabilistic model for LoS propagation was adopted in \cite{8254658}, which is, unfortunately, too complicated to yield tractable analysis.
On the other hand, there are few existing works for cellular-connected UAV networks in which the ground base stations (GBSs) of large-scale terrestrial cellular networks cause interference. Comprehensive analysis of coverage and rate was carried out in \cite{azari2018cellular}. Nevertheless, the obtained bulky analytical results, typically involving multiple nested integrals, do not provide insights for network design.

This paper proposes a new analytical framework for cellular-connected UAV networks, which leads to results that are more tractable than those provided by previous studies. We first introduce a new probabilistic model for LoS propagation in such networks, called the \emph{LoS ball} model, which not only achieves high accuracy but also enables tractable analysis. 
Subsequently, an effective approach to handle general Nakagami fading channels is proposed. 
With the help of the developed analytical framework, we evaluate
the coverage probability of downlink cellular-connected
UAV networks for a random spatial network model, where the
GBSs are modeled as a homogeneous Poisson point process
(PPP).
In particular,
we derive a novel and tractable expression for the coverage probability of cellular-connected UAV networks, which contains a single integral operation and can be efficiently computed.
Based on our analytical results, the impact of the GBS density is investigated. It is shown that there exists an optimal GBS density that maximizes the coverage probability. A lower bound on the optimal GBS density is accordingly derived, which is shown to be tight via numerical results. 
To the best of the authors' knowledge, this is the first analytical result on the impact of the GBS density in cellular-connected UAV networks.

\section{System Model}
In this section, we describe the random spatial network model, channel model, and blockage model for cellular-connected UAV networks.
\subsection{Network and Channel Models}\label{II-A}
\begin{figure}[t]
	\centering\includegraphics[width=6.8cm]{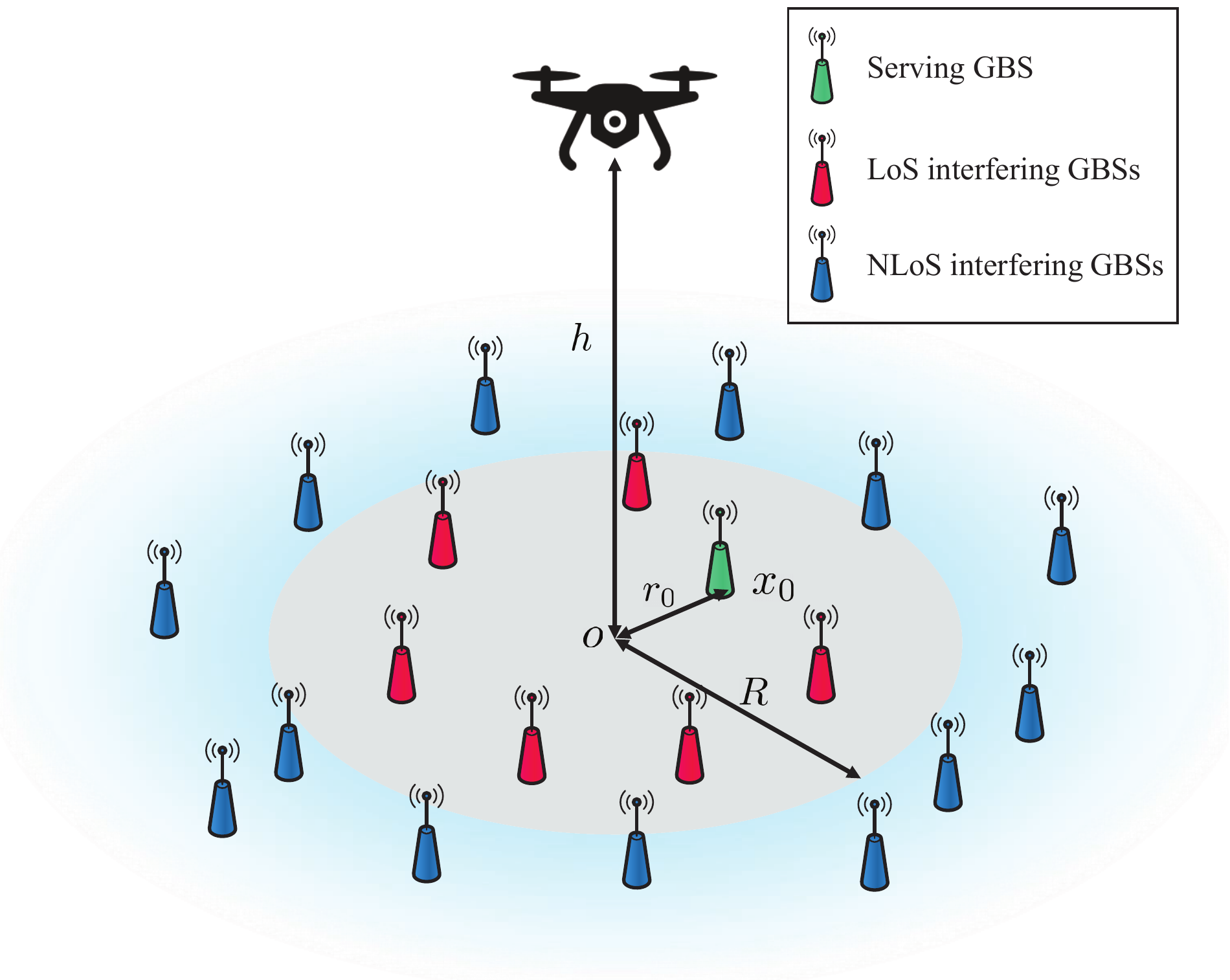}
	\caption{A sample network model is shown where GBSs are modeled as a homogeneous PPP, and an aerial user is associated with the nearest GBS. The LoS ball is used to model the blockages in cellular-connected UAV networks.}\label{systemmodel}
\end{figure}

Consider a downlink cellular-connected UAV network, where GBSs are distributed according to a homogeneous PPP $\Phi$ with density $\lambda$ \cite{haenggi2012stochastic}, as shown in Fig. \ref{systemmodel}.
We focus on the performance analysis of a typical aerial user located at altitude $h$, whose projection on the ground is at the origin $o$.
We assume that the typical aerial user is served by the nearest GBS located at $x_0\in\Phi$. All GBSs are assumed to  transmit with the same transmit power $P_\mathrm{t}$ using a single antenna; the extension to the multi-antenna case is discussed in Remark 1.
The received signal for the typical aerial user is given by
\begin{equation}\label{signal}
y=\sqrt{P_\mathrm{t}\zeta(r_0)}\bar{h}_{x_0}s_{x_0}+\sum_{x\in\Phi\backslash \{x_0\}}\hspace{-0.1em}\sqrt{P_\mathrm{t}\zeta\left(\left\Vert x\right\Vert\right)}\bar{h}_xs_x+n_{x_0},
\end{equation}
where $r_0$ is the ground distance from the associated GBS to the origin, i.e.,  $r_0=\left\Vert x_0\right\Vert$.
The path loss and small-scale fading from the GBS located at $x$ to the UAV are denoted by
$\zeta\left(\left\Vert x\right\Vert\right)$ and  $\bar{h}_x$, respectively. The symbol transmitted by the GBS at $x$ is denoted as $s_x$ with unit variance, while $n_{x_0}\sim\mathcal{N}(0,\sigma^2)$ stands for the additive white Gaussian noise with variance $\sigma^2$.

In order to model the wireless link between a GBS and the UAV, the LoS and NLoS components are considered separately. The path loss $\zeta\left(\left\Vert x\right\Vert\right)$ is given by
\begin{equation}
\zeta\left(\left\Vert x\right\Vert\right)=A_v\left(\left\Vert x\right\Vert^2+h^2\right)^{-\frac{\alpha_v}{2}},\quad v\in\{\mathrm{L},\mathrm{N}\},
\end{equation}
where $v\in\{\mathrm{L},\mathrm{N}\}$ represents the propagation conditions of LoS or NLoS links, respectively. The path loss exponent is denoted as $\alpha_v$, and $A_v$ is the path loss at the reference distance $\sqrt{\left\Vert x\right\Vert^2+h^2}=1$ m.

In this paper, we assume that the small-scale fading $\bar{h}_x$ is Nakagami-$M$ fading, where $M$ is the Nakagami fading parameter. This fading is widely assumed for modeling LoS components in wireless channels.  We assume an integer Nakagami fading parameter $M_\mathrm{L}>1$ for the LoS links. 
In conventional cellular networks where the communication links are mainly NLoS, the small-scale fading is assumed to be Rayleigh fading, i.e., $M_\mathrm{N}=1$. We also adopt this assumption in this paper and assume the NLoS propagation follows Rayleigh fading.
\subsection{LoS Ball Model}
In this subsection, we present the first ingredient of the proposed analytical framework, i.e., the LoS probability in the blockage model.
Based on measurement data, the LoS probability in UAV networks was given in \cite[eq. (4)]{azari2018cellular}, \cite[pp. 26]{instance1290}. However, these complicated expressions do not yield tractable analytical results and  may only be suitable for simulations. Therefore, a tractable LoS probability function that maintains accuracy is of great importance in the analysis of cellular-connected UAV networks.

The investigation of the LoS probability can be traced back to the studies of millimeter wave (mm-wave) networks, where the signals with higher frequencies are sensitive to blockages. A simple yet effective blockage model called the \emph{LoS ball} was proposed in \cite{6932503}, where the model was shown to be an accurate approximation for the exact one. More importantly, the LoS ball model was shown to lead to a tractable analysis of mm-wave networks \cite{6932503,7913628}. Inspired by these previous works, we resort to this
model and introduce it to cellular-connected UAV networks.

As shown in Fig. \ref{systemmodel}, we define an LoS radius $R$ in the LoS ball model, which represents the average distance between the projection of an aerial user and its nearby blockages. In this way, the LoS probability of a certain link is one when the GBS is located within radius $R$, and zero otherwise.
Compared with mm-wave networks, a key difference of the LoS ball model in cellular-connected UAV networks is that the model is not only 2-D distance-dependent, but also altitude-dependent in the 3-D space.
In principle, the LoS radius $R$ should be a monotonically increasing function of the UAV altitude $h$, denoted as $R(h)$. The higher the UAV flies, the more GBSs can be seen via an LoS path. 
The function $R(h)$ also depends on the environment (rural, urban, downtown with high-rises, and suburbs with simple houses, etc.). 
The analysis in this paper is valid for arbitrary functions $R(h)$, and the derivation of a specific expression for $R(h)$ is deferred to future work.

By adopting the LoS ball model, the interfering GBSs can be split into two categories. In particular, the set of interfering GBSs is denoted by $\Phi\backslash\{x_0\}=\Phi_\mathrm{L}\cup\Phi_\mathrm{N}$. As shown in Fig. \ref{systemmodel}, the LoS interfering GBSs form a PPP $\Phi_\mathrm{L}$ conditioned on $r_0$ in an annulus, with inner radius $r_0$ and outer radius $R$, and the NLoS interfering GBSs compose a PPP $\Phi_\mathrm{N}$ in $\mathbb{R}^2$ outside a disk with radius $R$.

\begin{figure}[t]
	\centering\includegraphics[width=6.8cm]{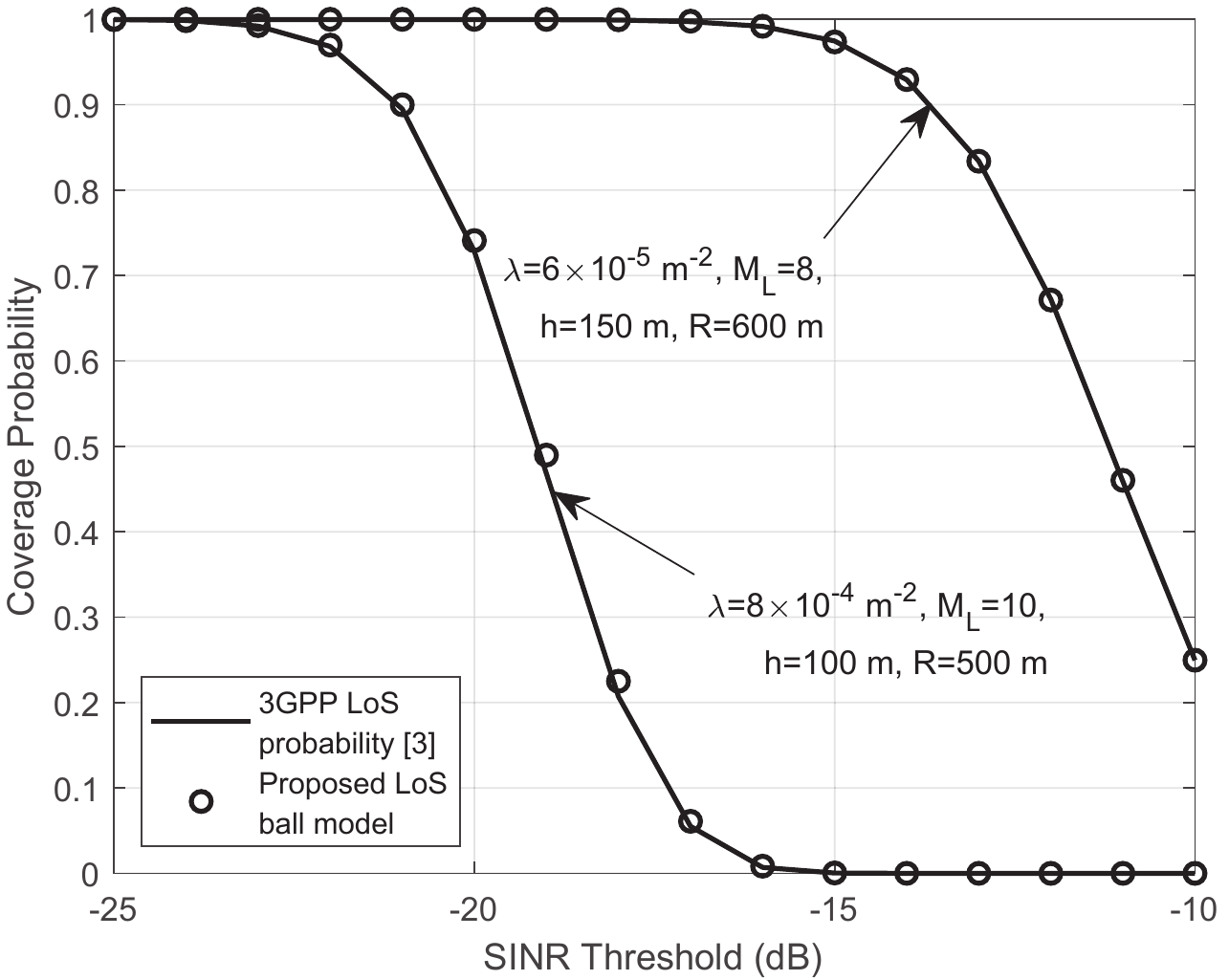}
	\caption{The coverage probability with the 3GPP LoS probability model and the proposed LoS ball model.}\label{validation}
\end{figure}

In Fig. \ref{validation}, we evaluate the coverage probability by adopting the 3GPP LoS probability \cite[p. 26]{instance1290} and the proposed LoS ball model, respectively. As can be observed, the proposed LoS ball model is an accurate approximation of the measurement-based model, and can effectively capture the probabilistic propagation in cellular-connected UAV networks.
Furthermore, we shall reveal the tractability of this model in the next section.

\section{Coverage Analysis of Cellular-Connected UAV Networks}\label{s3}
In this section, we first present an expression for the signal-to-interference-plus-noise ratio (SINR), and then derive a novel approach for coverage analysis based on the LoS ball model and general Nakagami fading channels.

\subsection{SINR}
According to \eqref{signal}, the receive SINR of a typical aerial user is given by
\begin{align}
\mathrm{SINR}&=\frac{P_\mathrm{t}\zeta(r_0)\left|\bar{h}_{x_0}\right|^2}
{\sigma^2+\displaystyle\sum_{x\in\Phi\backslash\{x_0\}}P_\mathrm{t}\zeta\left(\left\Vert x\right\Vert\right)\left|\bar{h}_x\right|^2}\label{SINR}\\
&\triangleq\frac{g_{x_0,v}\left(r_0^2+h^2\right)^{-\frac{\alpha_v}{2}}}
{\sigma_\mathrm{n}^2
+\sum_{v\in\{\mathrm{L},\mathrm{N}\}}\sum_{x\in\Phi_v}g_{x,v} \left(\left\Vert x\right\Vert^2+h^2\right)^{-\frac{\alpha_v}{2}}
},\nonumber
\end{align}
where $g_{x_0,v}=A_v\left|\bar{h}_{x_0}\right|^2$ is the signal power gain, $g_{x,v}=A_v\left|\bar{h}_{x}\right|^2$ is the interference power gain, and $\sigma_\mathrm{n}^2=\sigma^2/P_\mathrm{t}$ is the normalized noise power. 
Since, according to the assumptions in Section \ref{II-A}, the small-scale fading is assumed to be Nakagami fading, both the signal and interference power gains are independent and identically Gamma distributed as $\mathrm{Gamma}\left(M_v,A_v/M_v\right)$. Later we shall see that the general Nakagami fading model is the main obstacle for a tractable analysis.

\emph{Remark 1:} In Section \ref{II-A}, we assumed that each GBS is equipped with a single antenna. 
However, our analysis can be readily extended to the scenario where each GBS is equipped with multiple sectored antennas. As the GBSs are tilted downwards to the ground users, the UAV is typically served with sidelobe gains $G_\mathrm{s}$ \cite{azari2018cellular}. Correspondingly, both $g_{x_0,v}$ and $g_{x,v}$ are independent and identically Gamma distributed as $\mathrm{Gamma}\left(M_v,G_\mathrm{s}A_v/M_v\right)$. Note that the extension only affects the scale parameter of the Gamma distribution, and all the results derived in this paper still apply. Since the antenna tilting is not the main focus of this work, it is not included here to avoid an unnecessarily complicated presentation. 

\begin{figure*}
	\newcounter{TempEqCnt}                         			
	\setcounter{TempEqCnt}{\value{equation}} 			
	\setcounter{equation}{\value{longequ}}          	
	\begin{align}\label{longeq}
	c_n&= \frac{(-1)^{\delta_\mathrm{L}-n}\delta_\mathrm{L}\Gamma(M_\mathrm{L}+n)}{\Gamma(M_\mathrm{L})\Gamma(n+1)}
	\tau^{\delta_\mathrm{L}}  \left[B\left(-\tau;n-\delta_\mathrm{L},1-n-M_\mathrm{L}\right)- B\left(- \left(\frac{u+h^2}{R^2+h^2}\right)^{\frac{\alpha_\mathrm{L}}{2}}\tau;n-\delta_\mathrm{L},1-n-M_\mathrm{L}\right)\right](u+h^2)\nonumber\\
	&\relphantom{=}+{(-1)^{\delta_\mathrm{N}-n}\delta_\mathrm{N}}
	\left(\frac{\tau M_\mathrm{L}A_\mathrm{N}}{A_\mathrm{L}}\right)^{\delta_\mathrm{N}} B\left(- \frac{(u+h^2)^{\frac{\alpha_\mathrm{L}}{2}}}{(R^2+h^2)^{\frac{\alpha_\mathrm{N}}{2}}}\frac{\tau M_\mathrm{L}A_\mathrm{N}}{A_\mathrm{L}};n-\delta_\mathrm{N},-n\right)(u+h^2)^\frac{\alpha_\mathrm{L}}{\alpha_\mathrm{N}}
	\end{align}
	\hrule
\end{figure*}
\setcounter{equation}{\value{TempEqCnt}} 		

\subsection{Coverage Analysis}
As modern cellular networks are typically interference-limited, we focus on the signal-to-interference ratio (SIR) instead of the SINR.  We will later justify this assumption through simulations.
The \emph{coverage probability}, defined as the probability that the received SIR is greater than a threshold $\tau$, is written as
\begin{equation}\label{coverage}
p_\mathrm{c}=\mathbb{P}\left(\mathrm{SIR}>\tau\right)=P_\mathrm{L}p_\mathrm{c,\mathrm{L}}+P_\mathrm{N}p_\mathrm{c,\mathrm{N}},
\end{equation}
where $P_\mathrm{L}$ and $P_\mathrm{N}$ are the probabilities that the UAV is associated with an LoS and an NLoS GBS, respectively. According to the void probability of homogeneous PPPs \cite{haenggi2012stochastic}, these two probabilities are given by
\begin{equation}
P_\mathrm{L}=1-e^{-\pi\lambda R^2},\quad P_\mathrm{N}=1-P_\mathrm{L}=e^{-\pi\lambda R^2}.
\end{equation}
The remaining two terms in \eqref{coverage}, i.e., $p_\mathrm{c,\mathrm{L}}$ and $p_\mathrm{c,\mathrm{N}}$, are the coverage probabilities assuming that the UAV is associated with an LoS GBS and an NLoS GBS, respectively. According to \eqref{SINR} and \eqref{coverage}, these two terms can be expressed in a general form as
\begin{align}
p_{\mathrm{c},v}
&=\mathbb{P}\left[g_{x_0,v}>\tau \left(r_0^2+h^2\right)^{\frac{\alpha_v}{2}}I\right]\nonumber\\
&\overset{(a)}{=}\mathbb{E}_{r_0}\left\{\sum_{n=0}^{M_v-1}\frac{s^n}{n!}\mathbb{E}_I\left[I^ne^{-s I}\middle|r_0\right]\right\}\nonumber\\
&=\mathbb{E}_{r_0}\left[\sum_{n=0}^{M_v-1}\frac{(-s)^n}{n!}\mathcal{L}^{(n)}(s)\right]\triangleq\mathbb{E}_{r_0}\left[p_{\mathrm{c},v|r_0}\right],\label{nth}
\end{align}
where $I=\sum_{v\in\{\mathrm{L},\mathrm{N}\}}\sum_{x\in\Phi_v}g_{x,v} \left(\left\Vert x\right\Vert^2+h^2\right)^{-\frac{\alpha_v}{2}}$, $s=M_v\tau \left(r_0^2+h^2\right)^{\frac{\alpha_v}{2}}/A_v$, and $\mathcal{L}(s)=\mathbb{E}_I\left[e^{-sI}\middle|r_0\right]$ is the Laplace transform of interference conditioned on $r_0$.
The notation $\mathcal{L}^{(n)}(s)$ stands for the $n$-th derivative of $\mathcal{L}(s)$, and step $(a)$ involves the cumulative distribution function of Gamma random variable $g_{x_0}\sim\mathrm{Gamma}\left(M_v,A_v/M_v\right)$.

It is observed from \eqref{nth} that, due to the general Nakagami fading, the conditional coverage probabilities critically depend on the $n$-th derivatives of the Laplace transform $\mathcal{L}(s)$. This is the main obstacle for obtaining tractable analytical results. The derivatives were computed in a brute-force manner in previous works \cite{7967745,8254658,azari2018cellular}, which led to extremely tedious analytical results.
In contrast, we present the second ingredient of the proposed framework, i.e., a novel approach to handle general Nakagami fading.
The main steps of this effective approach are listed in Methodology 1, based on which tractable results for the conditional coverage probabilities $p_{\mathrm{c},v|r_0}$ are derived in the following.

\floatname{algorithm}{Methodology}
\begin{algorithm}[t]
	\label{alternating}
	\caption{Main Steps for Tractable Coverage Analysis}
	\begin{algorithmic}[1]
		\STATE 
		Derive the conditional log-Laplace transform $\eta_v(s)$ of the aggregate interference from $\Phi_v$;
		\STATE 
		Calculate the $n$-th ($1\le n\le M_v-1$) derivatives of $\eta_v(s)$ to populate the entries $t_n=\frac{(-s)^n}{n!}\eta_v^{(n)}(s)$ of the lower triangular Toeplitz matrix
		\begin{equation}
		\mathbf{T}_{M_v}=\begin{bmatrix}
		t_0&&&\\
		t_1&t_0&&\\
		\vdots&&\ddots&\\
		t_{M_\mathrm{L}-1}&\dots&t_1&t_0\\
		\end{bmatrix};
		\end{equation}
		\STATE
		Derive the conditional coverage probabilities according to $p_{\mathrm{c},v|r_0}=\left\Vert e^{\mathbf{T}_{M_v}}\right\Vert_1$, where $||\cdot||_1$ denotes the induced $\ell_1$-norm of a matrix.
		\STATE Calculate the coverage probability according to \eqref{coverage}.
	\end{algorithmic}
\end{algorithm}

\begin{thm}\label{th1}
	Based on the LoS ball model, the conditional coverage probability when the typical aerial user is associated with an LoS GBS is given by
	\begin{equation}\label{eq7}
	p_{\mathrm{c},\mathrm{L}|r_0}=e^{\pi\lambda u}\left\Vert e^{\pi\lambda\mathbf{C}_{M_\mathrm{L}}}\right\Vert_1.
	\end{equation}
	The non-zero entries of the lower triangular Toeplitz matrix $\mathbf{C}_{M_\mathrm{L}}$ are given by \eqref{longeq}, where $u=r_0^2$, $\delta_v=\frac{2}{\alpha_v}$, $\Gamma(\cdot)$ is the Gamma function, and $B(\cdot;\cdot,\cdot)$ is the incomplete Beta function.
\end{thm}
\begin{IEEEproof}
	A sketch of the proof is given in Appendix \ref{appA}.
\end{IEEEproof}
\setcounter{equation}{9} 
\emph{Remark 2:} Theorem \ref{th1} presents a novel representation of the conditional coverage probability, which is more compact than existing analytical results, and therefore enables efficient numerical evaluation. 
More importantly, in Section IV, we shall see that the new representation, along with the proposed LoS ball model, is able to reveal key network insights.
\begin{coro}
	Based on the LoS ball model, the conditional coverage probability when the typical aerial user is associated with an NLoS GBS is given by
	\begin{equation}
	p_{\mathrm{c},\mathrm{N}|r_0}=\exp\left\{\pi\lambda (u+h^2)\left[1-N_0\right]\right\},
	\end{equation}
	where $N_0=-\delta_\mathrm{N}(-\tau)^{\delta_\mathrm{N}}B\left(-\tau;-\delta_\mathrm{N},0\right)$.
\end{coro}
\begin{IEEEproof}
The UAV is associated with an NLoS GBS, which indicates that all the interfering GBSs are in NLoS conditions. Furthermore, Rayleigh fading is assumed for NLoS links, i.e., $M_\mathrm{N}=1$. Therefore, the corollary is a reduced scalar version of Theorem 1 with slight modifications, which completes the proof.
\end{IEEEproof}

The probability density functions of the serving GBS distance $r_0$ assuming that the typical aerial user is associated with an LoS and an NLoS GBS are given by
\begin{equation}
f_{r_0|\mathrm{L}}(r)=2\pi\lambda re^{-\pi\lambda r^2}/P_\mathrm{L},\quad 0\le r\le R,
\end{equation}
and 
\begin{equation}
f_{r_0|\mathrm{N}}(r)=2\pi\lambda re^{-\pi\lambda r^2}/P_\mathrm{N},\quad r\ge R,
\end{equation}
respectively, with which we can complete the evaluation of the coverage probability in \eqref{coverage}.
\begin{prop}\label{main}
Based on the LoS ball model, the SIR coverage probability $p_\mathrm{c}$ in \eqref{coverage} is given by
\begin{equation}
p_\mathrm{c}= \pi\lambda e^{\pi\lambda h^2}\int_{0}^{R^2}\left\Vert e^{\pi\lambda\mathbf{C}_{M_\mathrm{L}}(u)}\right\Vert_1\mathrm{d}u+\frac{e^{\pi\lambda \left[h^2- N_0(h^2+R^2)\right]}}{N_0}.\label{eq13}
\end{equation}
\end{prop}
\begin{IEEEproof}
	The proof is completed by substituting the results in Theorem 1 and Corollary 1 into \eqref{coverage}.
\end{IEEEproof}

\emph{Remark 3:} In existing works \cite{7412759,7967745,8254658,azari2018cellular}, the coverage probabilities were derived either as recursive expressions or complicated forms with multiple nested integrals, sums, and products. In contrast, the coefficients $c_n(u)$ in Theorem \ref{th1} can be expressed in closed form based on the incomplete Beta function, and the single integral involved in the coverage probability in Proposition \ref{main} can be efficiently calculated with modern numerical software. This underlines the tractability of the proposed framework, which is further used to disclose a key network insight in Section IV.

\begin{figure}[t]
	\centering\includegraphics[width=6.8cm]{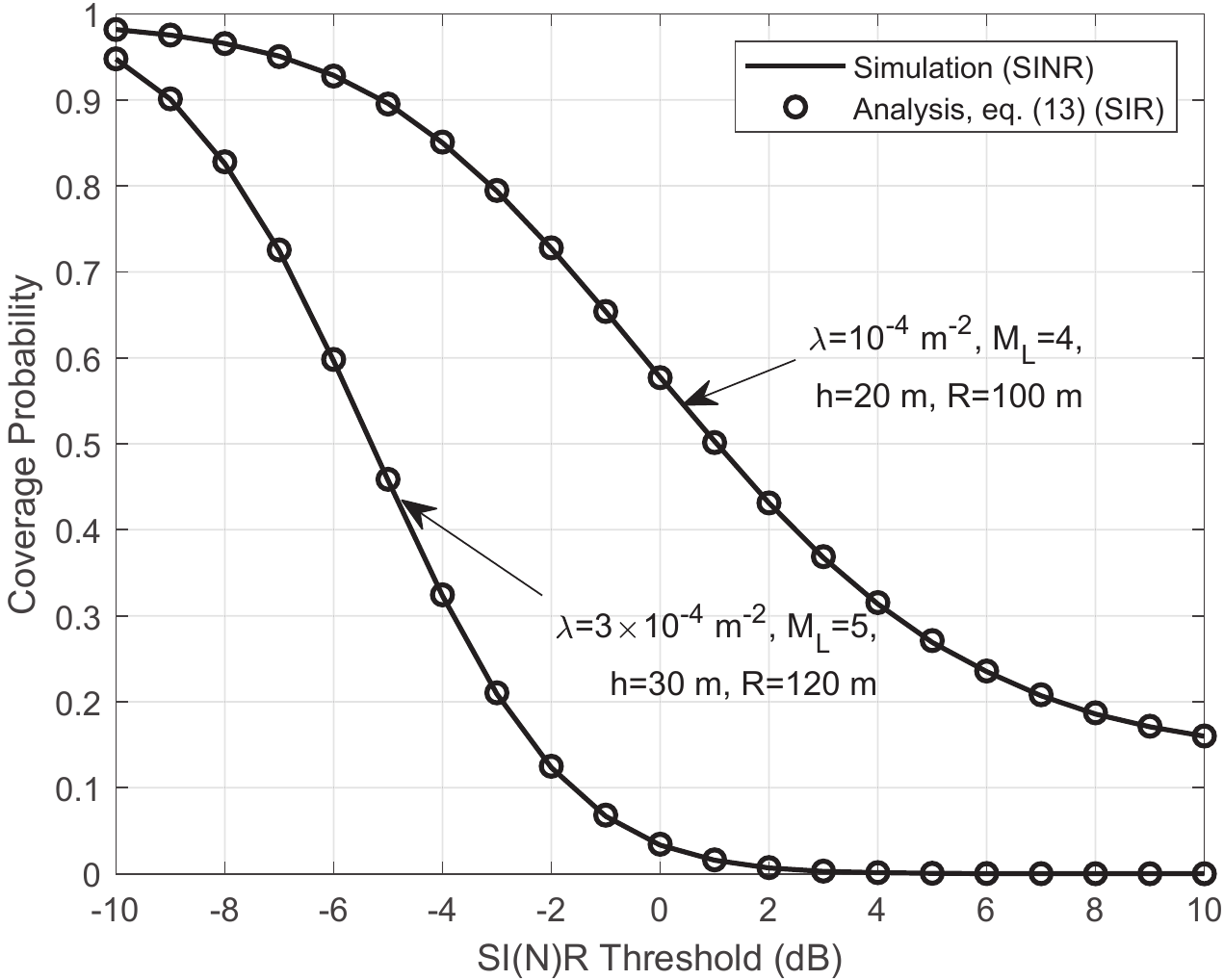}
	\caption{The SI(N)R coverage probability of cellular-connected UAV networks when $\sigma_\mathrm{n}^2=−97$ dBm, $\alpha_\mathrm{L}=2.1$, $\alpha_\mathrm{N}=4$,  $A_\mathrm{L}=-41.1$ dB, and $A_\mathrm{N}=-32.9$ dB.}\label{fig1}
\end{figure}

In Fig. \ref{fig1}, we plot the SINR coverage probabilities using simulation and compare them with the analytical results for the SIR coverage derived in Proposition \ref{main}. As can be observed, the numerical results based on \eqref{eq13} are in excellent agreement with the simulation results, which means that the influence of the noise is negligible and the interference-limited assumption is justified.
Furthermore, the good match of these two results also indicates the accuracy of our derived analytical results.

\section{Impact of the GBS Density}
In conventional cellular networks without considering the LoS probability, there is a well-known property called the SIR invariance in terms of the base station (BS) density. Specifically, the coverage probability is invariant to the BS density \cite{6042301}. This was extended to the multi-slope path loss model \cite{7061455}, for which the coverage probability was proved to be a monotonically decreasing function of the BS density. However, these conclusions no longer hold in cellular-connected UAV networks, where the LoS probability and different small-scale fading models are considered. The following result reveals the effect of the GBS density on the coverage probability in cellular-connected UAV networks.
\begin{prop}\label{prop3}
	Based on the LoS ball model, there exists an optimal GBS density $\lambda^\star$ that maximizes the coverage probability, and a lower bound on the optimal GBS density $\lambda_\mathrm{LB}\le\lambda^\star$
	is one of the solutions to the following polynomial equation
	\begin{equation}\label{poly}
	\sum_{n=0}^{M_\mathrm{L}}\beta_n\lambda^n=0.
	\end{equation}
	The coefficients of the polynomial are given by
	\begin{equation}
	\beta_n=\begin{cases}
	\pi R^2&n=0\\
	\int_0^{R^2}a(u)\kappa_{n-1}(u)\mathrm{d}u&n=M_\mathrm{L}\\
	\int_0^{R^2}a(u)\kappa_{n-1}(u)+(n+1)\kappa_n(u)\mathrm{d}u&\text{otherwise},
	\end{cases}
	\end{equation}
	where $a(u)=\pi\left(c_0(u)+h^2\right)$, and 
	\begin{equation}
		 \kappa_n(u)=\frac{\pi^{n+1}\left\Vert\left(\mathbf{C}_{M_\mathrm{L}}(u)-c_0(u)\mathbf{I}_{M_\mathrm{L}}\right)^n\right\Vert_1}{n!}.
	\end{equation}
\end{prop}
\begin{IEEEproof}
	A sketch of the proof is given in Appendix \ref{appB}.
\end{IEEEproof}

\emph{Remark 4:} With this result, for a given network setting, e.g., the path loss exponent, UAV altitude, and Nakagami parameters, the coefficients of the polynomial can be efficiently computed via numerical integration, and therefore the lower bound on the optimal GBS density can be obtained by solving \eqref{poly}. Note that abundant efficient methods are available for solving polynomial equations \cite{sturmfels2002solving}, and equation \eqref{poly} can be solved in closed form up to order $M_\mathrm{L}=4$ according to the Abel-Ruffini theorem. Thanks to the LoS ball model and the compact form derived in Theorem \ref{th1}, we are able to analyze the impact of GBS density, which cannot be unraveled by existing works \cite{7412759,7967745,8254658,azari2018cellular}.

\begin{figure}[t]
	\centering\includegraphics[width=6.8cm]{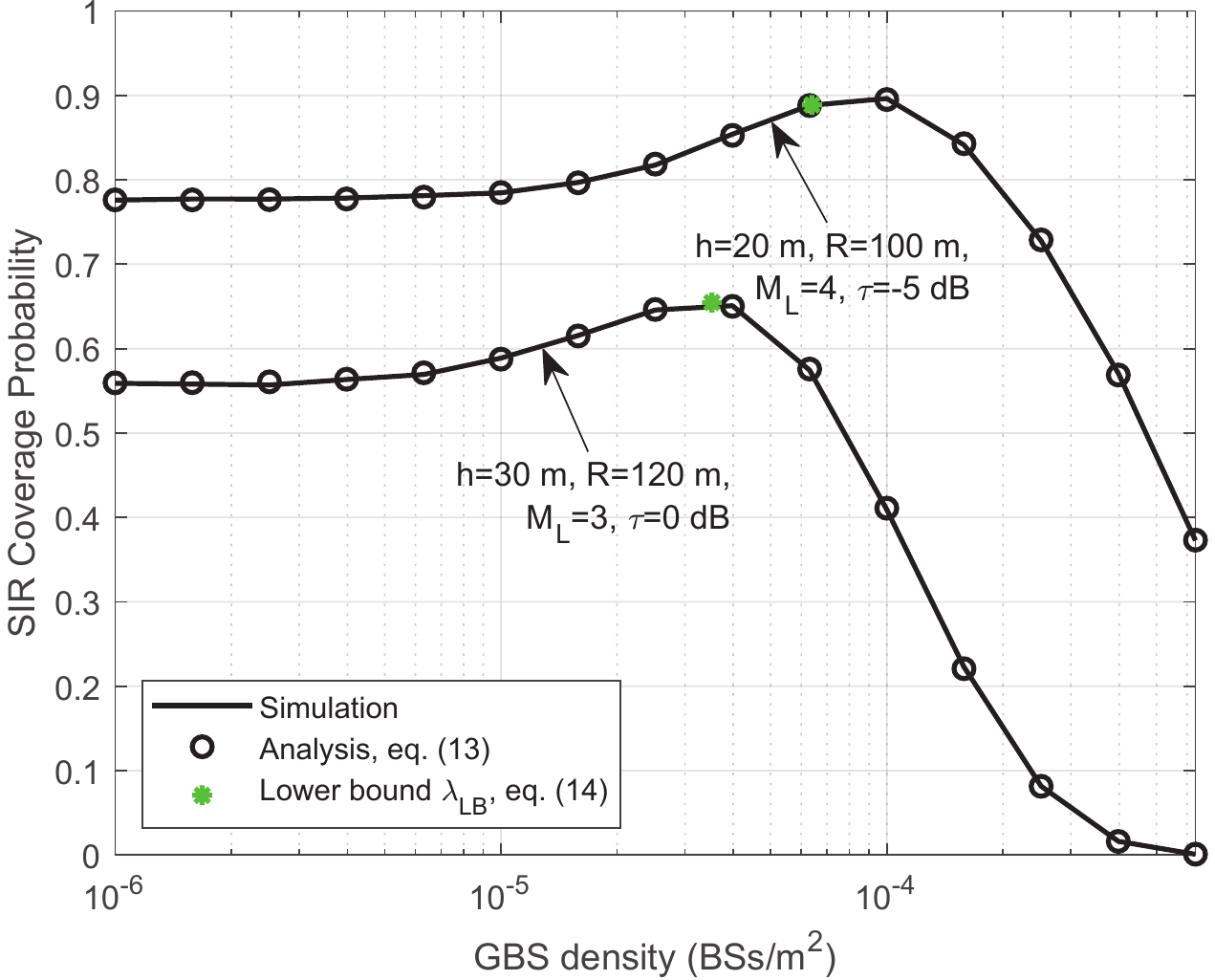}
	\caption{The SIR coverage probability of cellular-connected UAV networks when $\alpha_\mathrm{L}=2.1$, $\alpha_\mathrm{N}=4$,  $A_\mathrm{L}=-41.1$ dB, and $A_\mathrm{N}=-32.9$ dB.}\label{fig2}
\end{figure}

In Fig. \ref{fig2}, the SIR coverage probability is plotted versus the GBS density. As can be observed, there exists a peak value for the coverage probability, and the lower bound $\lambda_\mathrm{LB}$ is close to the optimal density $\lambda^\star$, which shows the effectiveness of the results in Proposition \ref{prop3}. Note that searching for the optimal GBS density $\lambda^\star$ via simulation is computationally heavy and the accuracy depends on the search step size. In contrast, closed-form expressions for $\lambda_\mathrm{LB}$  can be directly obtained from Proposition \ref{prop3}.

The phenomenon that the coverage probability has a peak value does not occur for conventional wireless networks as studied in \cite{6042301,7061455}.
This is because of the difference in the small-scale fading for LoS and NLoS propagation, which were previously assumed to be the same in \cite{7061455}. 
When the GBS density gradually increases, the signal link tends to be LoS with a higher probability, and therefore experiences Nakagami fading rather than Rayleigh fading. This change in small-scale fading results in a slight increase of the SIR coverage probability, which also implicitly indicates that Nakagami fading provides better coverage than Rayleigh fading.
When the GBS density further increases, there are more and more interfering LoS GBSs that cause severe interference, which decreases the SIR coverage probability.
Hence, the coverage probability is maximized when the signal link is LoS with Nakagami fading while keeping most of the interfering GBSs in NLoS conditions with Rayleigh fading.
A similar phenomenon was numerically found for mm-wave networks \cite{7913628}. More insights for network design can be revealed in terms of the UAV altitude, which is a key differentiating feature in cellular-connected UAV networks.
Recall that the LoS radius $R$ is a monotonically increasing function of the UAV altitude. When the UAV altitude drops, the LoS radius reduces, and therefore denser deployment of the GBSs is required to allow for an LoS signal link to maximize the coverage probability, which is confirmed in Fig. \ref{fig2}. Furthermore, mainly due to the shorter signal link distance, the maximum coverage probability increases when the UAV descends nearer to the ground.

\section{Conclusions}
By introducing an LoS ball model to the coverage analysis of cellular-connected UAV networks, this paper developed an analytical framework for the coverage probability that is more tractable than existing ones. In particular, it was demonstrated that the LoS ball model is an excellent candidate for tractable analysis of UAV networks, while maintaining satisfactory accuracy. 
More importantly, based on the derived tractable expression, it was discovered that there exists an optimal GBS density that maximizes the coverage probability, and a tight lower bound on the optimal GBS density was analytically derived.
More generally, it was shown that the modeling and analytical approaches applicable to cellular-connected UAV and mm-wave networks have many similarities.
In future work, it will be interesting to further refine the LoS ball model, e.g., by specifying the relation between the LoS radius and the UAV altitude, and apply it to conduct a more detailed analysis of cellular-connected UAV networks.
\appendices

\section{}\label{appA}
According to the probability generating functional of a PPP, the Laplace transform $\mathcal{L}_\mathrm{N}(s)$ of the interference in $\Phi_\mathrm{N}$ in \eqref{nth} is expressed as $\mathcal{L}_\mathrm{N}(s)=e^{\eta_\mathrm{N}(s)}$, where $\eta_\mathrm{N}(s)$ is given by
\begin{align}
&\relphantom{=}-2\pi\lambda\int_R^\infty{\left(1-\mathbb{E}_{g}[\exp(-sg(r^2+h^2)^{-\frac{\alpha_\mathrm{N}}{2}})]\right)}r\mathrm{d}r\nonumber\\
&=-\pi\lambda\int_{R^2+h^2}^\infty{\left(1-\mathbb{E}_{g}[\exp(-sgr^{-\frac{\alpha_\mathrm{N}}{2}})]\right)}\mathrm{d}r\nonumber\\
&\overset{(b)}{=}
\pi\lambda (R^2+h^2)+\pi\lambda(R^2+h^2)
\delta_\mathrm{N} \left[s(R^2+h^2)^{-\frac{\alpha_\mathrm{N}}{2}}\right]^{\delta_\mathrm{N}}\nonumber\\
&\relphantom{=}\times\mathbb{E}_g\left[g^\delta\gamma(-\delta_\mathrm{N},s(R^2+h^2)^{-\frac{\alpha_\mathrm{N}}{2}}g)\right],
\end{align}
where $\gamma(\cdot,\cdot)$ is the lower incomplete Gamma function, and $g$ is distributed as $\mathrm{Gamma}\left(M_\mathrm{N},A_\mathrm{N}/M_\mathrm{N}\right)$. Step $(b)$ is derived from \cite[eq. (4)]{6042301} by change of variables in the integral.

It has been shown in our previous work \cite{8490204} that the conditional coverage probability $p_{\mathrm{c},\mathrm{L}}$ in Theorem \ref{th1} is critically determined by the log-Laplace transform $\eta(s)$, and can be expressed as in \eqref{eq7}. The non-zero entries of the lower triangular Toeplitz matrix $\mathbf{C}_{M_\mathrm{L}}$ are given by
\begin{align}
&\relphantom{=}\frac{(-s)^n}{n!}\eta^{(n)}(s)\nonumber\\
&\overset{(c)}{=} (R^2+h^2)\frac{(-1)^{\delta_\mathrm{N}-n}\delta_\mathrm{N}\Gamma(M_\mathrm{N}+n)}{\Gamma(M_\mathrm{N})\Gamma(n+1)}\frac{r_0^2+h^2}{R^2+h^2}
\left(\frac{\tau M_\mathrm{L}A_\mathrm{N}}{M_\mathrm{N}A_\mathrm{L}}\right)^{\delta_\mathrm{N}} \nonumber\\
&\relphantom{=}\times \pi\lambda B\left(- \frac{\left(r_0^2+h^2\right)^{\frac{\alpha_\mathrm{L}}{2}}}{\left(R^2+h^2\right)^{\frac{\alpha_\mathrm{N}}{2}}}\frac{\tau M_\mathrm{L}A_\mathrm{N}}{M_\mathrm{N}A_\mathrm{L}};n-\delta,1-n-M_\mathrm{N}\right),\nonumber\\
&\relphantom{=}+\mathds{1}(n=0)\pi\lambda (R^2+h^2).\label{eq18}
\end{align}
where $\mathds{1}(\cdot)$ denotes the indicator function, and step $(c)$ uses a similar derivation method as \cite[Appendix A]{8490204}.

Similarly, the log-Laplace transform of the interference in $\Phi_\mathrm{L}$ conditioned on $r_0$ is given by
\begin{equation}
-2\pi\lambda\int_{r_0}^R{\left(1-\mathbb{E}_{g}[\exp(-sg(r^2+h^2)^{-\frac{\alpha_\mathrm{L}}{2}})]\right)}r\mathrm{d}r,
\end{equation}
and the following steps to derive the non-zero coefficients of the lower triangular Toeplitz matrix are derived in the same manner as in \eqref{eq18}, which completes the proof.

\section{}\label{appB}
By applying the derivation steps in \cite[eq. (72)]{8490204},
the integrand of the first term in \eqref{eq13} is rewritten as
\begin{equation}
\pi\lambda e^{\pi\lambda h^2}\left\Vert e^{\pi\lambda\mathbf{C}_{M_\mathrm{L}}(u)}\right\Vert_1=
e^{a(u)\lambda}
\sum_{n=0}^{M-1}\kappa_n(u)\lambda^{n+1}.
\end{equation}
Therefore, a lower bound on the derivative of the coverage probability with respect to the GBS density is given by
\begin{equation}
p^\prime_\mathrm{c}(\lambda)=\frac{\partial p_\mathrm{c}}{\partial \lambda}\ge\pi R^2+\int_0^{R^2}e^{a(u)}I(u)\mathrm{d}u\overset{(d)}{\ge}\sum_{n=0}^{M_\mathrm{L}}\beta_n\lambda^n,\\
\end{equation}
where $I(u)=\sum_{n=1}^{M_\mathrm{L}-1}\left[a(u)\kappa_{n-1}(u)+(n+1)\kappa_n(u)\right]\lambda^n+a(u)\kappa_{M_\mathrm{L}-1}(u)\lambda^{M_\mathrm{L}}$, and step $(d)$ applies the Chebyshev integral inequality. 
Note that the function $p_\mathrm{c}^\prime(\lambda)$ has only one zero and is monotonically decreasing in the neighborhood of the zero. Therefore, the zero of a lower bound on the function $p_\mathrm{c}^\prime(\lambda)$ is a lower bound on the zero of the function $p_\mathrm{c}^\prime(\lambda)$.
By setting the lower bound on the derivative to zero, the proof is completed.

\bibliographystyle{IEEEtran}
%
\bibliography{bare_conf}

\end{document}